\documentclass[twocolumn]{aastex63}

\usepackage{amssymb, amsfonts, amsmath,framed}

\usepackage{bm}
\usepackage{graphicx}
\usepackage{epstopdf,color,verbatim}
\usepackage{threeparttable}
\usepackage{amsmath}
\usepackage{amsmath}	
\usepackage{amssymb}	
\usepackage[]{txfonts}

\expandafter\ifx\csname package@font\endcsname\relax\else
 \expandafter\expandafter
 \expandafter\usepackage
 \expandafter\expandafter
 \expandafter{\csname package@font\endcsname}
\fi
\expandafter\ifx\csname package@font\endcsname\relax\else
 \expandafter\expandafter
 \expandafter\usepackage
 \expandafter\expandafter
 \expandafter{\csname package@font\endcsname}
\fi
\newcommand{\figg}[1]{Fig.~\ref{fig:#1}}
\newcommand{\eqq}[1]{Eq.~\ref{eq:#1}}
\def\bi{\begin{itemize}}
\def\ei{\end{itemize}}
\def\bq{\begin{equation}}
\def\eq{\end{equation}}
\def\bqy{\begin{eqnarray}}
\def\eqy{\end{eqnarray}}
\newcommand{\rhot}{r_{\rm L,hot}}

\newcommand{\be}{\begin{equation}}
\newcommand{\ee}{\end{equation}}

\newcommand{\unit}[1]{\,{\rm #1}}

\newcommand{\comp}{c/\omega_{\rm p}}
\newcommand{\omp}{\omega_{\rm p}}

\begin{document}
\title{Global Kinetic Modeling of the Intrabinary Shock in Spider Pulsars}

\author[0000-0001-5186-6195]{Jorge Cort\'es}
\email{jorgecortes@astro.columbia.edu}
\affiliation{Department of Astronomy and Columbia Astrophysics Laboratory, Columbia University, New York, NY 10027, USA}

\author[0000-0002-1227-2754]{Lorenzo Sironi}
\email{lsironi@astro.columbia.edu}
\affiliation{Department of Astronomy and Columbia Astrophysics Laboratory, Columbia University, New York, NY 10027, USA}

\begin{abstract}
Spider pulsars are compact binary systems composed of a millisecond pulsar and a low-mass companion. The relativistic magnetically-dominated  pulsar wind impacts onto the companion, ablating it and slowly consuming its atmosphere. The interaction forms an intrabinary shock, a proposed site of particle acceleration. We perform global fully-kinetic particle-in-cell simulations of the intrabinary shock, assuming that the pulsar wind consists of plane-parallel stripes of alternating polarity and that the shock wraps around the companion. We find that particles are efficiently accelerated via shock-driven reconnection. We extract first-principles synchrotron spectra and lightcurves which are in good agreement with X-ray observations: (1) the synchrotron spectrum is nearly flat, $F_\nu\propto {\rm const}$; (2) when the pulsar spin axis is nearly aligned with the orbital angular momentum, the light curve displays two peaks, just before and after the pulsar eclipse (pulsar superior conjunction), separated in phase by  $\sim 0.8\, {\rm rad}$; (3) the peak flux exceeds the one at inferior conjunction by a factor of ten. We demonstrate that the double-peaked signature in the lightcurve is due to Doppler boosting in the post-shock flow.
\end{abstract}



\section{Introduction}
Millisecond pulsars (MSPs) are the product of a ``recycling process" that leads to their millisecond spin periods
through mass accretion from a binary companion \citep{alpar_1982}. MSPs differ from ``normal" pulsars by having weaker surface magnetic fields ($\mathrm{B_P}\sim10^8-10^9\,\unit{G}$), faster spin periods ($\mathrm{P\sim10^{-3}}\,\unit{s}$), smaller spin down rates ($\mathrm{\dot{P}}\sim10^{-20}\,\unit{s\,s^{-1}}$), and a higher probability of being in a binary system \citep{manchester_2017}.
An interesting subclass of MSPs are the so-called spider pulsars, the Redbacks (RBs) and Black Widows (BWs). RBs and BWs are in tight binary orbits (orbital period ${P_{\rm orb}} \lesssim 1\unit{day}$; orbital separation ${a} \sim 10^{11}\unit{cm}$) and possess either a non-degenerate companion with mass of ${\sim}\,0.1{-}0.5\,M_\odot$ (for RBs) or a degenerate companion with mass of ${\sim}\,0.01{-}0.05\,M_\odot$ (for BWs). In these systems, the pulsar possesses a spin-down luminosity of $\dot{E}_{\rm SD} \sim 10^{34 } - 10^{35}$ erg s$^{-1}$, emanating an intense relativistic wind. When impacting onto the companion star, the wind exacerbates ablation and mass loss of the companion \citep{phinney_1988,ginzburg_quataert_2020}, which is then  ``devoured'' by the pulsar wind --- hence, the evocative name of these systems.

The MSP relativistic wind consists of toroidal stripes of opposite magnetic field polarity, separated by current sheets of hot plasma \citep{bogo_99}. For obliquely-rotating pulsars, this is the configuration expected around the equatorial plane of the wind, where the sign of the toroidal field alternates with the pulsar period \citep{petri_lyubarsky_2007}. The wind terminates in a strong shock, where the ram pressure of the pulsar wind is balanced by the companion wind or its magnetosphere. The intrabinary shock may be an efficient site of particle acceleration \citep{harding_gaisser_1990, arons_tavani_1993}, which would then explain why X-ray lightcurves of spider pulsars are often modulated on the orbital period \citep[e.g.,][]{bogdanov_grindlay_vandenberg_2005,bogdanov_2011,bogdanov_21,huang_2012,roberts_2014,roberts_2015}. The observed X-ray flux, which originates from synchrotron emission of relativistic electrons and positrons at the intrabinary shock, peaks just before and after the
pulsar eclipse, which has been attributed to Doppler boosting effects caused by
the post-shock flow \citep{romani_sanchez_2016,sanchez_romani_2017,wadiasingh_2017,wadiasingh_2018,kandel_romani_an_2019,vandermerwe_2020}. The X-ray spectrum is markedly nonthermal, and has flat photon indices $\Gamma_X \approx1-1.5$ \citep{roberts_2015} that imply rather hard underlying electron distributions with a slope of $p=-d \log N/d\log\gamma= 2 \Gamma_X -1 \approx 1-2$.

Relativistic hydrodynamic and magnetohydrodynamic (MHD) calculations \citep{bogo_08,bogo_12,bogo_19,boschramon_2012,boschramon_barkov_perucho_2015,lamberts_2013,huber_2021} allow to determine the overall morphology and dynamics of the system. Yet, they cannot probe the properties of the non-thermal particles responsible for the emission. In this paper, we present the first global kinetic simulations of the intrabinary shock in spider pulsars,  assuming that the shock wraps around the companion star. We employ large-scale particle-in-cell (PIC) simulations, which allow to capture at the same time the global shock morphology and the microphysics of particle acceleration. We extract first-principles synchrotron spectra and lightcurves, and show that they are in good agreement with X-ray observations. We demonstrate that the observed double-peaked signature in the X-ray lightcurve is due to Doppler boosting in the post-shock flow.

\section{Length Scales in Spider Pulsars}
 The pulsar emits a relativistic magnetically-dominated wind composed of stripes of alternating field polarity. Decay of the Poynting-flux-dominated stripes occurs at a radius of $R_{\rm diss} \sim 10^2-10^4\,R_{\rm LC}$, where $R_{\rm LC}=c/\Omega=2\pi c P$ is the pulsar light cylinder radius \citep{cerutti_philippov_2017,cerutti_philippov_dubus_2020}. For the Crab, the termination shock is at $R_{\rm TS} \sim 10^9\,R_{\rm LC}\gg R_{\rm diss}$. In contrast, in spider pulsars the intrabinary shock is at $R_{\rm TS}\sim a\sim10^{10} - 10^{11}\,\mathrm{cm}$, corresponding to $R_{\rm TS} \sim 10^2 - 10^4\,R_{\rm LC}$ for the typical light cylinder radii of MSPs ($R_{\rm LC}\simeq3\times10^7\, (\Omega/10^3\,{\rm s^{-1}})^{-1}\,\mathrm{cm}$). It follows that the pulsar wind arrives at the intrabinary shock still in the form of magnetically-dominated stripes. The compression of the flow at the shock can drive stripe annihilation via shock-driven magnetic reconnection, resulting in efficient particle acceleration \citep{sironi_spitkovsky_2011,lu_2021}. 

We now characterize the hierarchy of length scales in spider pulsars. We consider systems in which the pulsar wind is more powerful than the companion wind, so the shock wraps around the companion star, as inferred, e.g., for the original BW PSR B1957+20. The radius of curvature of the shock $R_{\rm curv}$ is then comparable to the size $R_{\ast}\sim 10^{10}\unit{cm}$ of the companion star \citep{burrows_1993}. Since the distance between the pulsar and the shock is $R_{\rm TS}\sim a\gtrsim R_{\rm curv}$, we assume for simplicity that the pulsar wind can be modeled as a sequence of plane-parallel stripes. The ratio of  shock curvature radius to the stripe wavelength $\lambda=2\pi R_{\rm LC}$ is
\begin{equation} \label{eq:r_lam}
\vspace{-0.0cm}
    \frac{R_{\rm curv}}{\lambda}\sim 5\times 10^1\left ( \frac{R_{\rm curv}}{10^{10}\unit{cm}} \right ) \left ( \frac{\Omega}{10^3\unit{s^{-1}}} \right )~.
    \vspace{-0.cm}
\end{equation}
 The ratio of stripe wavelength to the typical post-shock Larmor radius $\rhot$ is \citep{sironi_spitkovsky_2011}
\begin{equation} \label{eq:lam_comp}
    \frac{\lambda}{\rhot}\sim 4\pi\kappa \frac{R_{\rm LC}}{R_{\rm TS}} \sim 3\times 10^1 \left ( \frac{\kappa}{10^4} \right ) \left ( \frac{10^{11}\,\mathrm{cm}}{{R_{\rm TS}}} \right ) \left ( \frac{10^3\,\mathrm{s^{-1}}}{\Omega} \right )
\end{equation}
assuming a wind multiplicity \citep{goldreich_julian_1969} of $\kappa \sim 10^4$ \citep{timokin_12, timokhin_harding_2015, philippov_timokhin_spitkovsky_2020}. Local PIC simulations of shock-driven reconnection at the termination shock of striped winds have demonstrated that for $\lambda/\rhot\gg1$ the particle spectrum approaches a broad power-law tail of index $1<p<2$ \citep{sironi_spitkovsky_2011}, as required to explain the X-ray spectra of spider pulsars.


\section{Simulation Setup}
\label{sec:sims}
We use the 3D electromagnetic PIC code TRISTAN-MP \citep{buneman_1993, spitkovsky_2005}. We employ a 2D spatial domain in the $x-y$ plane, but we track all three components of velocity and  electromagnetic fields. The magnetically-dominated electron-positron pulsar wind propagates along $-\hat{x}$. It is injected from a moving boundary, that starts just to the right of the companion and moves along $+\hat{x}$ at the speed of light $c$.
This strategy  allows to save memory and computing time, while retaining all the regions that are causally connected with the initial setup of the system \citep{sironi_spitkovsky_2011,sironi_13}.  An absorbing layer for particles and fields is placed at $x=0$ (leftmost side of the domain). Periodic boundaries are used along the $y$ direction.
The pulsar wind magnetic field is initialized as
\begin{equation}
    B_y(x,t) = B_0\,\mathrm{tanh} \left\{ \frac{1}{\Delta} \left[ \alpha + \mathrm{cos} \left( \frac{2\pi(x+\beta_0ct}{\lambda} \right) \right] \right\}
    \label{eq:BB}
\end{equation}
where $\beta_0=(1-1/\gamma_0^2)^{1/2}$ is the  wind velocity and $\gamma_0$ the bulk Lorentz factor. We present results for $\gamma_0=3$, but we have verified that a choice of $\gamma_0=10$ leads to the same conclusions, apart from an overall shift in the energy scale \citep[see also][]{sironi_spitkovsky_2011}. The field direction flips across current sheets of thickness $\sim \Delta\lambda$. The field strength $B_0$ is parameterized via the magnetization $\sigma \equiv B_0^2/4 \pi \gamma_0 m n_{\rm 0} c^2=10$ (i.e., the ratio of Poynting to kinetic energy flux). Here, $m$ is the electron (or positron) mass and $n_{\rm 0}$ the density of particles in the ``cold wind'' (i.e. the region outside of current sheets, which instead have peak density of $4\,n_{\rm 0}$). 
Finally, $\alpha$ is a measure of the magnetic field averaged over one wavelength, such that $\langle B_y\rangle_\lambda/B_0=\alpha/(2-|\alpha|)$. A value of $\alpha=0$ --- or equivalently, ``positive'' and ``negative'' stripes of comparable width --- indicates the equatorial plane of the pulsar, whereas $|\alpha|$ increases when moving away from the midplane \citep{sironi_spitkovsky_2011}. For BWs, the spin axis of the pulsar is believed to be well aligned with the orbital angular momentum \citep{roberts_2013}, so we will only explore small values of $\alpha$, from 0 up to 0.3. 

\begin{figure*}
    \begin{center}
        \includegraphics[width=\textwidth, angle=0]{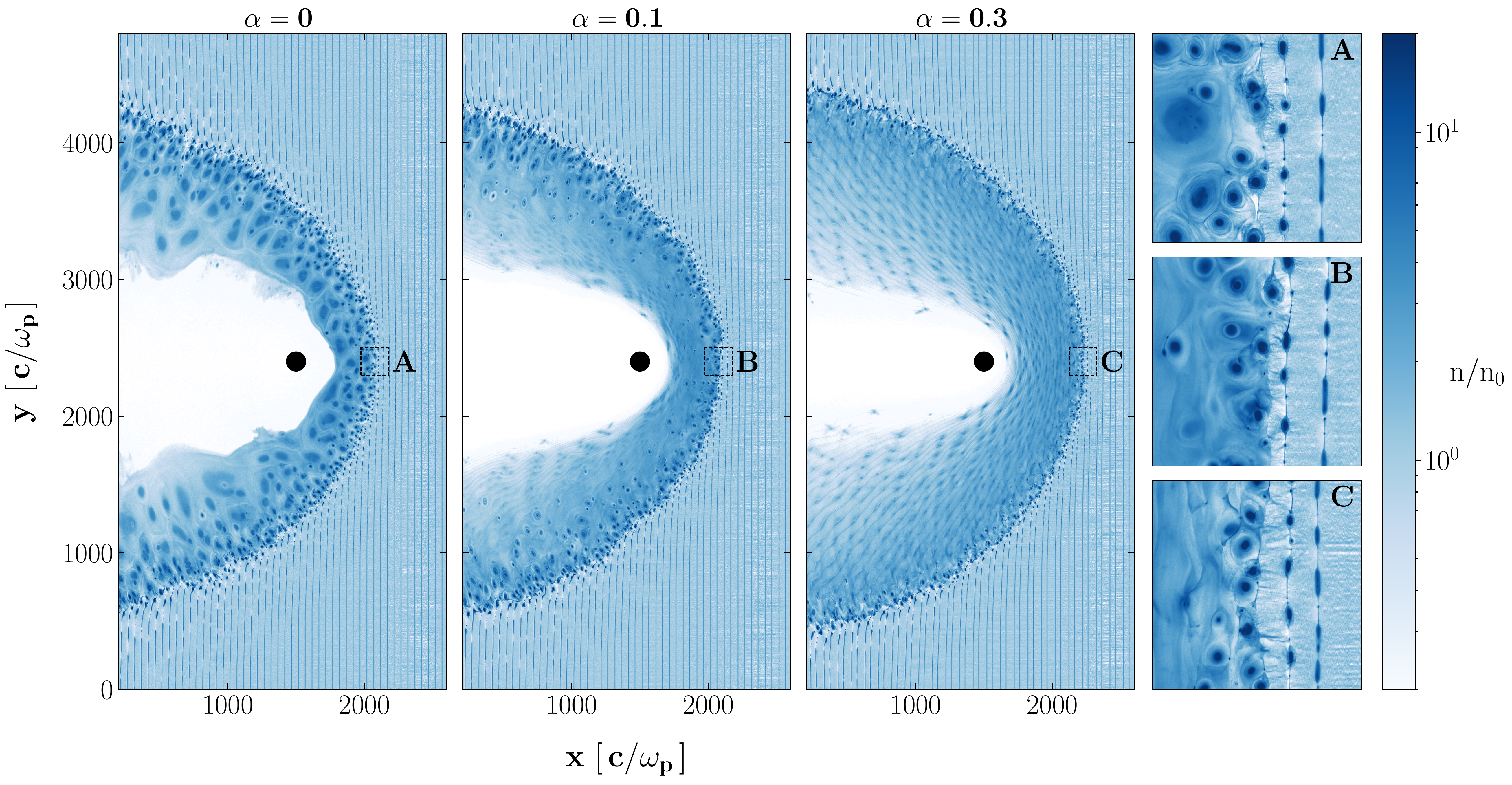}
        \caption{2D plots of the  number density  of pulsar wind particles, in units of $n_0$, for $\alpha=0, 0.1$ and $0.3$ (left to right panels). The plots refer to $\omp t = 3510$. The black filled circle represents the companion star. Insets A, B, and C --- marked in the main panels by a square box delimited by black dashed lines --- provide a zoom-in view  highlighting the formation of magnetic islands / plasmoids due to shock-driven reconnection.}
        \label{fig:dens}
    \end{center}
\end{figure*}

\begin{figure*}
    \begin{center}
        \includegraphics[width=\textwidth, angle=0]{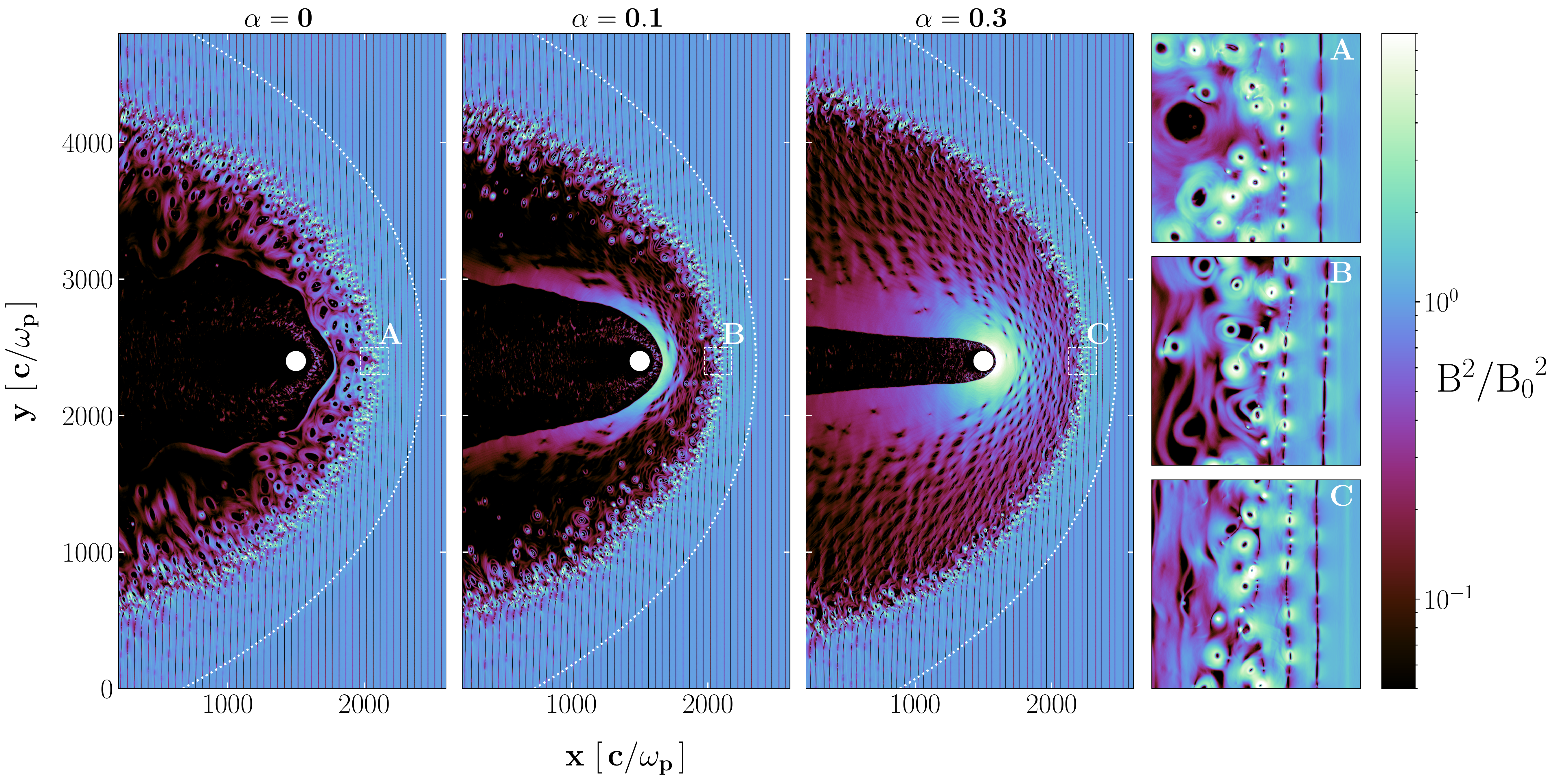}
        \caption{2D plots of the magnetic energy density, in units of the upstream value $B_0^2/8\pi$,  for $\alpha=0, 0.1$ and $0.3$ (left to right). The plots refer to $\omp t = 3510$. The white filled circle represents the companion star. Insets A, B, and C --- marked in the main panels by a square box delimited by white dashed lines --- are as in \figg{dens}. Dotted white lines indicate the fast MHD shock.}
        \label{fig:eB}
    \end{center}
\end{figure*}

The relativistic skin depth in the cold wind $\comp \equiv (\gamma_0 m c^2 / 4 \pi e^2 n_{\rm 0})^{1/2}$ is resolved with 10 cells, so the pre-shock Larmor radius $r_{\rm L} \equiv \gamma_0 m c^2 / e B_0=(\comp)/\sqrt{\sigma}$ is resolved with 3 cells. The post-shock Larmor radius, assuming complete field dissipation, is defined as $\rhot=\sigma r_{\rm L}$. The numerical speed of light is 0.45 cells/timestep. Within the cold wind, each computational cell is initialized with two pairs of cold ($kT/mc^2 = 10^{-4}$) electrons and positrons, but we have checked that higher numbers of particles per cell lead to the same results. The temperature in the over-dense current sheets is set by pressure balance.

Our computational domain is $4800\,\comp$ wide (in the $y$ direction), or equivalently 48,000 cells wide. The center of the companion is placed at $(x,y)=(1500,2400)\,\comp$, with a companion radius of $R_{\ast}=70\,\comp$. The companion surface (a cylinder, for our 2D geometry) is a conducting boundary for fields and a reflecting boundary for particles. The value for $R_{\ast}$ is chosen such that the companion wind (see below) is stopped by the pulsar wind at $R_{\rm curv} \simeq 200\,\comp$, which then gives the characteristic shock curvature radius. The large width of our domain in the $y$ direction is required such that the shock wraps around the companion until the $x=0$ open boundary, i.e., the shock surface does not cross the periodic $y$ boundaries. We set the stripe wavelength to be $\lambda=100\,\comp$. It follows that $R_{\rm curv}/\lambda\simeq 2$ and $\lambda/\rhot\simeq 30$, i.e., similar values (within one order of magnitude) as inferred in spider pulsar systems. Even larger values of $R_{\rm curv}/\lambda$ will be discussed in an upcoming work.

In our setup, the pulsar wind is stopped by a companion wind launched isotropically from its surface. We do not aim to reproduce the realistic properties of the companion wind. In fact, it would be numerically unfeasible to resolve the plasma scales in a dense non-relativistic companion wind, and yet perform global simulations of the system. Given that our focus is on pulsar wind particles, their acceleration and emission, the companion wind will merely serve to halt the pulsar wind. We initialize a companion wind with realistic values of the radial momentum flux, but with artificially smaller particle density (and so, artificially higher wind velocity) to make the problem computationally tractable. The momentum flux density of the companion wind at the companion surface is twice larger than the momentum flux of the pulsar wind. We have tested that different choices for the companion wind density (at fixed momentum flux) lead to identical results for the structure, dynamics and particle acceleration properties of the pulsar wind termination shock. In the remainder of this work we will only consider acceleration and emission of pulsar wind particles; therefore, by ``wind'' we will refer to the pulsar wind, unless otherwise specified.


\section{Flow Dynamics}
Figures \ref{fig:dens} and \ref{fig:eB} show the  number density of pulsar wind particles (\figg{dens}) and the magnetic energy density (\figg{eB}) on the global scale of the intrabinary shock.  From left to right, we show $\alpha=0$, 0.1 and 0.3. All panels refer to $\omp t = 3510$, when for $\alpha=0$ the system has approached a quasi-steady state. 
The interaction between the pulsar wind and the companion wind is expected to generate two shocks (a reverse shock into the companion wind, and a forward shock into the pulsar wind) and a contact discontinuity. The reverse shock is visible only for $\alpha=0$ and 0.1 in \figg{eB}, as an arc of small-scale magnetic fields just to the right of the companion. In our setup, the companion wind is unmagnetized, and the reverse shock is mediated by the filamentation (or, Weibel) instability, which creates intense small-scale fields \citep[e.g.,][]{sironi_13}. The contact discontinuity is clearly seen, as the surface wrapping around the companion that delimits the white region in \figg{dens} and  the dark/black region in \figg{eB}. Both pulsar wind particles and strong fields are primarily confined to the right of the contact discontinuity.

The forward shock that stops the pulsar wind is the location where the ordered plane-parallel structure of the pre-shock wind transitions towards a more disordered, turbulent medium that flows around the companion and ultimately exits the left boundary. As described in local PIC simulations of relativistic striped winds \citep{sironi_spitkovsky_2011,lu_2021}, at the shock the flow compresses and the alternating fields annihilate via shock-driven reconnection. In fact, the overdense, magnetized, quasi-circular blobs seen in the rightmost columns of both figures are the characteristic magnetic islands / plasmoids produced by reconnection. 

More precisely, reconnection starts ahead of the main shock, at a fast MHD shock (dotted white lines in \figg{eB}) that   propagates into the pre-shock striped flow, compressing the incoming current sheets and initiating reconnection. When reconnection islands grow so big to occupy the entire region between neighboring current sheets, the striped structure is erased, and the main shock forms (at the boundary where $B^2/B_0^2$ turns from cyan to purple in \figg{eB}). For $\alpha\leq 0.3$, the average post-shock particle energy is larger than in the pre-shock wind by a factor of $\simeq\sigma$. Equivalently, most of the pre-shock field energy has been transferred to the particles. 

For $\alpha\neq 0$, the stripe-averaged field $\langle B_y\rangle_\lambda/B_0=\alpha/(2-|\alpha|)$ does not annihilate, and its field lines secularly accumulate, for our 2D geometry, in front of the companion star (see \figg{eB} for $\alpha=0.1$ and 0.3).\footnote{The larger magnetic pressure accumulating around the companion may also be the reason why the contact discontinuity wraps more tightly around the companion for higher $\alpha$.} This accumulation of magnetic energy/pressure is expected to be alleviated in 3D, as the field lines can pass above/below the spherical companion (in contrast to the cylindrical companion in 2D).  

The plasmoids are primarily concentrated near the shock. For $\alpha=0$, it is apparent that they merge with each other and grow in scale while flowing away from the shock, with the largest plasmoids having a size that is comparable to global scales (e.g., to the shock curvature radius).

\newpage
\section{Particle Acceleration and Emission}
\begin{figure}
    \begin{center}
        \includegraphics[width=\columnwidth, angle=0]{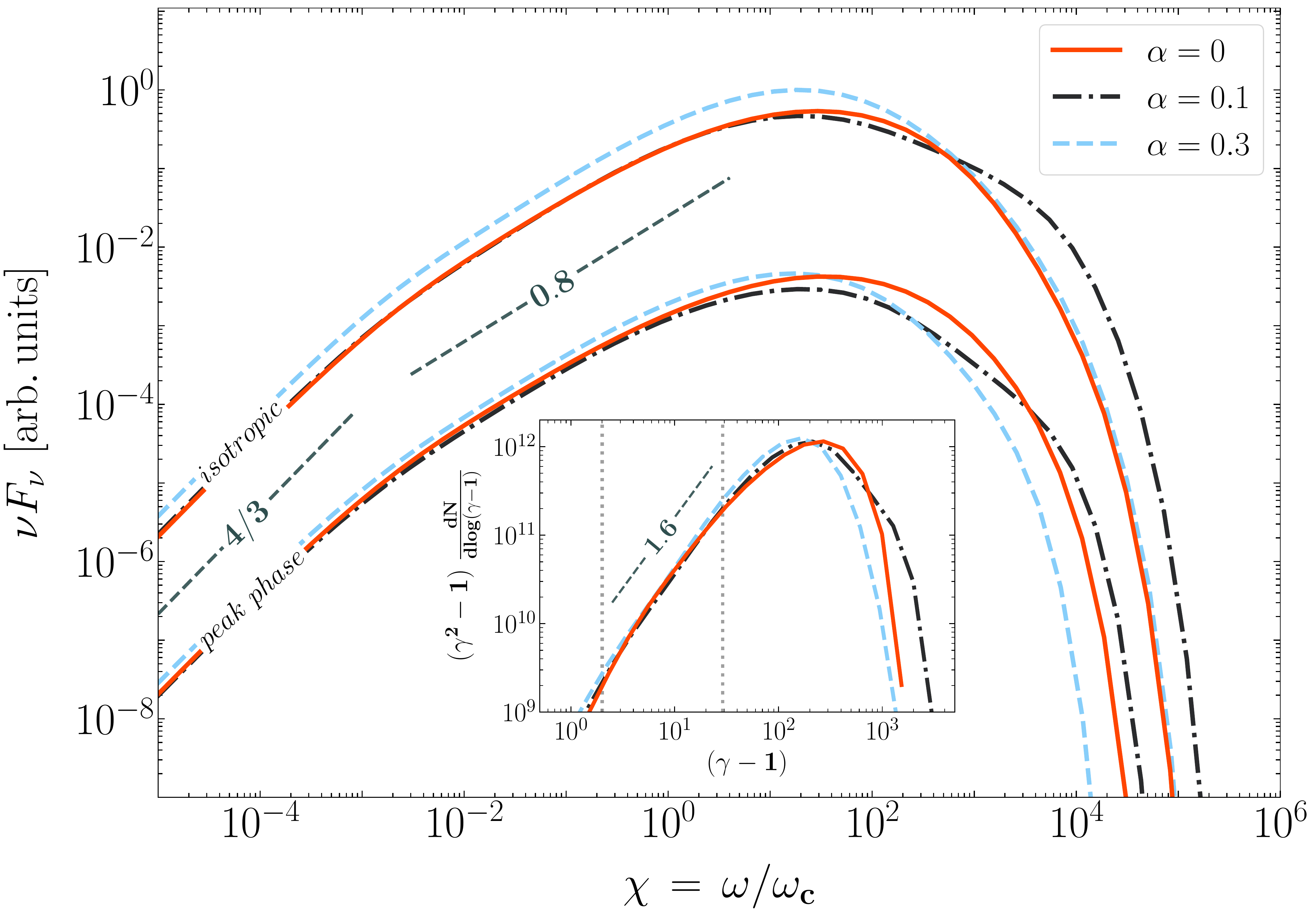}
        \caption{Synchrotron spectra $\nu F_\nu$ for $\alpha = 0$ (red solid), 0.1 (black dot-dashed) and $0.3$ (cyan dashed) at $\omp t = 3510$, the same time as in \figg{dens}. We define $\chi=\omega/\omega_{\rm c}$ where the characteristic synchrotron frequency $\omega_{\rm c}=(\gamma_0\sigma)^2 eB_0/mc$ is calculated for the average post-shock Lorentz factor $\gamma=\gamma_0\sigma$. We show both angle-integrated spectra (top series of lines) and spectra measured at the orbital phase $|\phi|= \pi/8$ corresponding to the two lightcurve peaks for $\alpha=0$ (``peak-phase'' spectra; bottom series). The subpanel shows the particle energy spectra for the different $\alpha$ cases (same color and linestyle coding as in the main panel). We multiply the particle count $dN/d\log(\gamma-1)$ by an additional factor of $(\gamma^2-1)$, proportional to the synchrotron power per particle. Vertical dotted lines in the subpanel are for $\gamma=\gamma_0$ and $\gamma=\gamma_0\sigma$.} 
        \label{fig:syncspectra}
    \end{center}
\end{figure}

\begin{figure}
    \begin{center}
        \includegraphics[width=\columnwidth, angle=0]{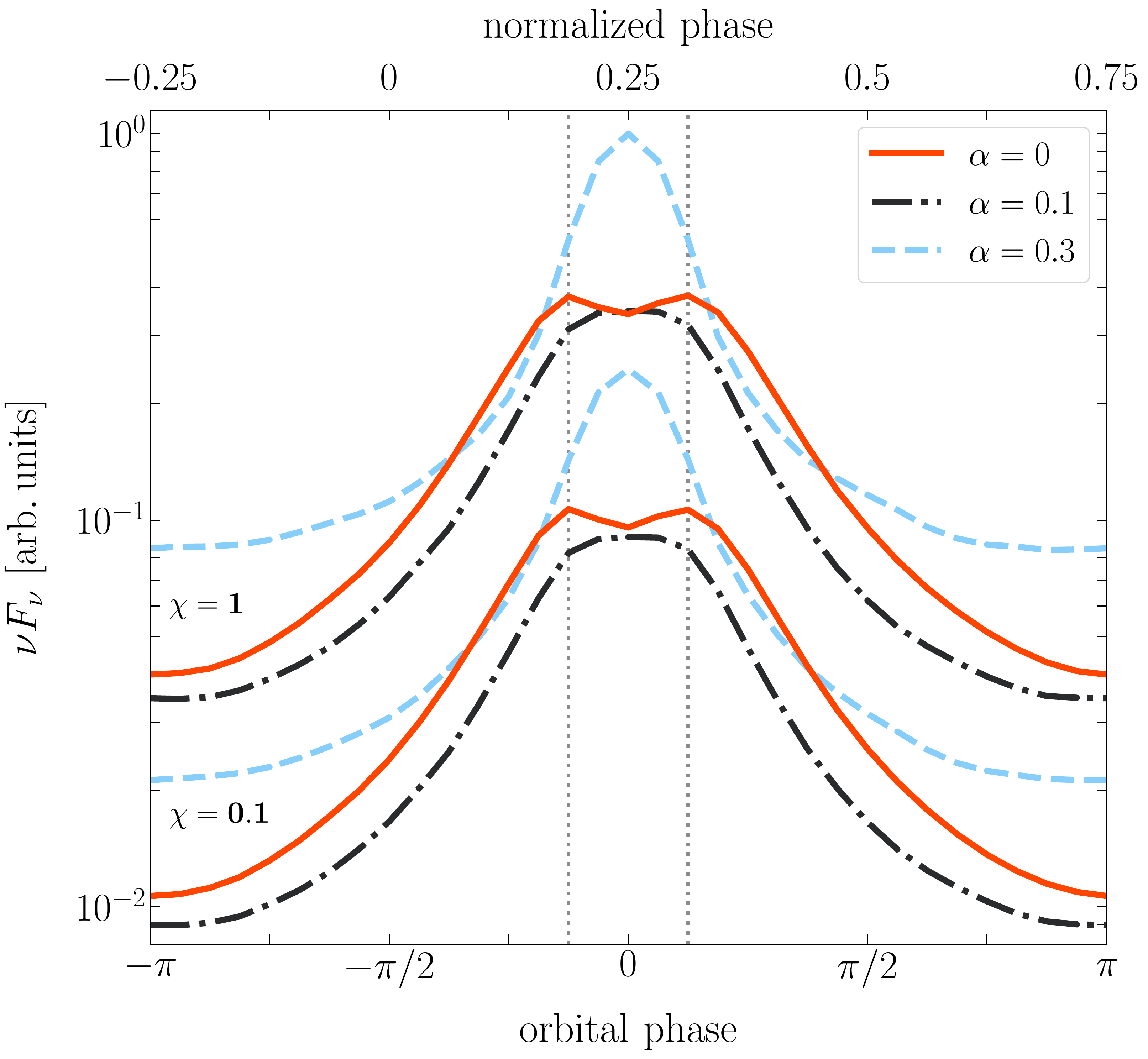}
        \caption{Synchrotron lightcurves for $\alpha=0$ (red solid), 0.1 (black dot-dashed) and $0.3$ (cyan dashed) at $\omp t = 3510$, as a function of orbital phase $\phi$ (bottom axis). We choose $\phi=0$ to correspond to superior conjunction (when the pulsar is eclipsed), while $\phi=\pm\pi$ to inferior conjunction. The top axis is the normalized phase, with 0.25 at the pulsar superior conjunction, as typical for X-ray observations. The lightcurves are shown for both  $\chi=0.1$ (bottom lines) and $1$ (top lines), where $\chi=\omega/\omega_{\rm c}$.
 Gray dotted lines at $|\phi|= \pi/8$ indicate the location of the two peaks for the $\alpha=0$ case.} 
        \label{fig:syncLC}
    \end{center}
\end{figure}

PIC simulations provide a first-principles assessment of the physics of particle acceleration and the properties of the resulting emission. We compute the synchrotron emission contributed by the post-shock pulsar wind particles.\footnote{Radiative cooling losses are not accounted for in the particle equation of motion, i.e., we assume a slow-cooling regime.} In Fig.~\ref{fig:syncspectra}, the top three curves show the angle-integrated synchrotron spectra for the three values of $\alpha$ explored in this work (see legend). The three lines with lower normalization, instead,  are the spectra measured at the orbital phase $|\phi|= \pi/8$ corresponding to the two lightcurve peaks for $\alpha=0$, as we show in \figg{syncLC} (``peak-phase'' spectra). Phase-resolved spectra are computed as follows. We consider a set of $N_{\rm los}=32$ lines of sight, all lying in the simulation plane (equivalently, the orbital plane) and equally spaced in angle between $-\pi$ and $\pi$. We take $\phi=0$ at superior conjunction (when the pulsar is eclipsed by the companion), and $\phi=\pm\pi$ at inferior conjunction. For each line of sight, the synchrotron emission is contributed by those particles whose velocity lies within an angle $\pi/N_{\rm los}$ from the line of sight. For each line of sight, we exclude the contribution from the spatial regions eclipsed by the companion (but we assume its wind to be optically thin).

All spectra  in \figg{syncspectra} display the usual $\nu F_\nu \propto \nu^{4/3}$ at low frequencies, with a transition to  $\nu F_\nu \propto  \nu^{0.8}$ in the range $\chi=\omega/\omega_{\rm c}\sim 10^{-2}-10^1$. A spectrum $\nu F_\nu \propto  \nu^{0.8}$ is in good agreement  with X-ray observations, since it corresponds to a photon spectral index $\Gamma_X\simeq 1.2$. Here, the characteristic synchrotron frequency $\omega_{\rm c}=(\gamma_0 \sigma)^2 eB_0/mc$ is calculated for the average post-shock Lorentz factor $\gamma=\gamma_0\sigma$ assuming efficient dissipation of the alternating stripes. Since $\sigma=10$ in our simulations, a value of $\chi\sim 10^{-2}$ then corresponds to the characteristic emission frequency of particles with $\gamma=\gamma_0$.

Further insight is provided by the downstream energy spectra of pulsar wind particles (subpanel in \figg{syncspectra}). There, we multiply the particle count $dN/d\log(\gamma-1)$ by an additional factor of $(\gamma^2-1)$, proportional to the synchrotron power per particle. Between $\gamma=\gamma_0=3$ and $\gamma=\gamma_0\sigma=30$, the particle spectrum can be modeled as a hard power law $dN/d\gamma\propto \gamma^{-p}$ with slope $p\simeq 1.4$. Local simulations of relativistic striped winds \citep{sironi_spitkovsky_2011} have shown that shock-driven reconnection leads, in the limit $\lambda/\rhot\gg1$, to hard power law distributions, with a slope consistent to what we measure here. It follows that the synchrotron spectral range between $\chi\sim 10^{-2}$ and $\chi\sim 10^1$ is contributed by particles accelerated via shock-driven reconnection. For $p\simeq 1.4$, the resulting synchrotron spectrum will be $\nu F_\nu \propto \nu^{(3-p)/2} = \nu^{0.8}$, as indeed observed in \figg{syncspectra}. The three $\alpha$ cases all exhibit the same spectral slope between $\chi\sim 10^{-2}$ and $\chi\sim 10^1$. Their difference at higher ($\chi\gtrsim 10^2$) frequencies mirrors the trend in the corresponding particle energy spectra, and it is due to differences in the acceleration physics of particles with $\gamma\gg \gamma_0\sigma$, as we will discuss in an upcoming work.

\begin{figure*}
    \begin{center}
        \includegraphics[width=\textwidth, angle=0]{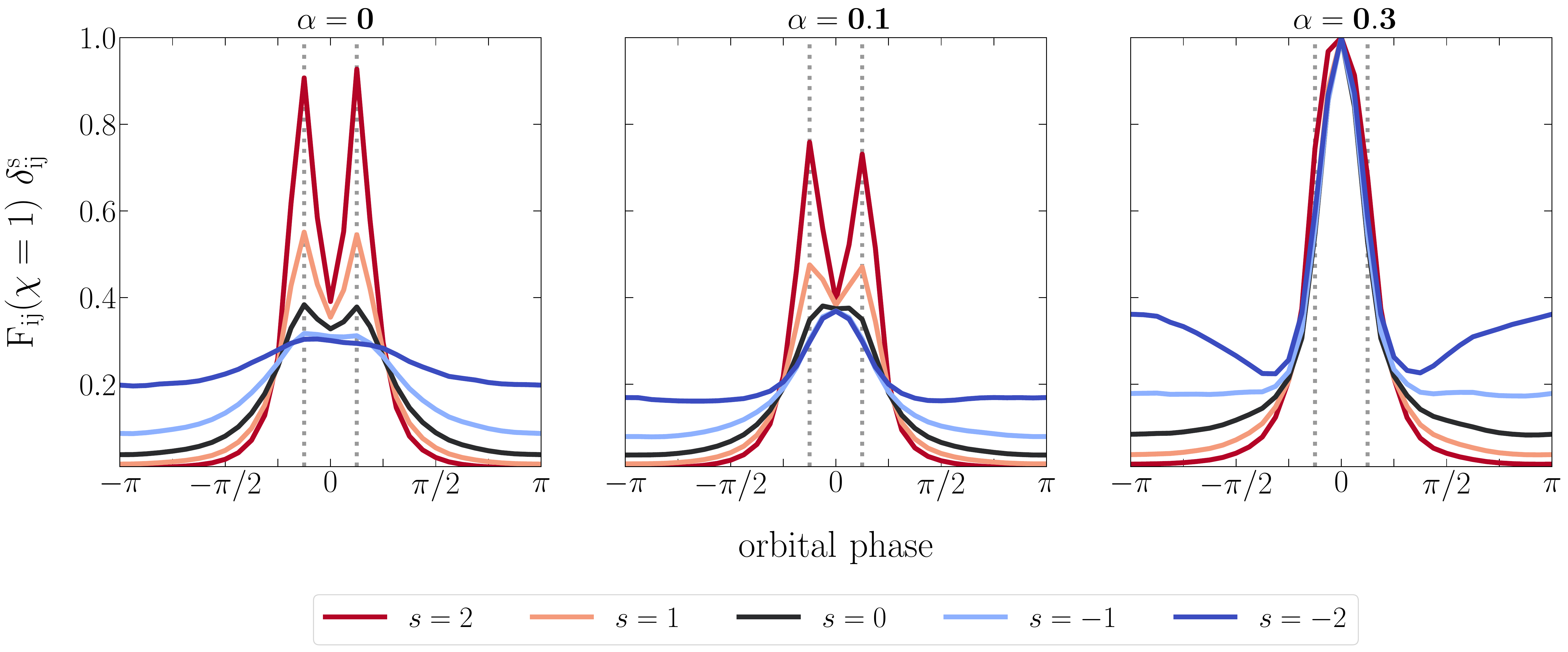}
        \caption{Synchrotron lightcurves for $\chi=1$ at $\omp t = 3510$, for different $\alpha=0$, 0.1 and 0.3 (from left to right). In every patch of the domain (of $100\times100$ cells), we weigh the corresponding synchrotron flux $F_{ij}(\chi=1)$ by different powers $s$ of the local Doppler factor $\delta_{ij}$, as indicated by the legend, and then we compute the sum $\Sigma_{i,j}F_{ij}(\chi=1)\delta_{ij}^s$. For each $s$, the lightcurves are normalized to the peak value of the corresponding $\alpha=0.3$ lightcurve.
        Gray dotted lines placed at $\phi=\pm \pi/8$ indicate the location of the two peaks for the $\alpha=0$ case.} 
        \label{fig:doppler}
    \end{center}
\end{figure*}

Phase-resolved lightcurves are shown in Figure \ref{fig:syncLC}, for two values of $\chi$ within the $\nu F_\nu\propto \nu^{0.8}$ spectral range: $\chi=1$ (top three lines) and $\chi=10^{-1}$ (bottom three lines). All curves are normalized to the peak of the $\alpha=0.3$ lightcurve at $\chi=1$. The two sets of lines display similar properties (they differ, at zeroth order, only by an overall normalization factor): the $\alpha=0.3$ lightcurve (cyan dashed) is strongly peaked at superior conjunction ($\phi=0$, when the pulsar is eclipsed by the companion); the peak is shallower for $\alpha=0.1$ (black dot-dashed); interestingly, for $\alpha=0$ (red solid) the lightcurve peaks at $\phi\simeq\pm\pi/8$, i.e., just before and after superior conjunction. For $\alpha=0$, the two peaks are separated by $\simeq 0.8\,$rad (equivalently, $\sim 0.13$ in normalized phase), and the flux at $\phi=0$ drops by $\sim 10\%$ below the peak flux. The flux at inferior conjunction ($\phi=\pm\pi$) is nearly a factor of 10 lower than the peak. All these properties hold throughout the spectral range $\chi=10^{-2}-10^1$ where $\nu F_\nu\propto \nu^{0.8}$, and are in good agreement with X-ray data \citep[e.g.,][]{bogdanov_grindlay_vandenberg_2005,bogdanov_2011,bogdanov_21,huang_2012,roberts_2014,roberts_2015}. 

Orbital modulations in the X-ray band have been attributed to Doppler boosting in the post-shock flow \citep{romani_sanchez_2016,wadiasingh_2017}. Our PIC simulations self-consistently include Doppler effects. Still, we can check the importance of Doppler boosting, by artificially over-emphasizing or de-emphasizing it. In \figg{doppler}, we divide the simulation domain in patches of $100\times100$ cells. In each patch, we compute the mean bulk fluid velocity $\boldsymbol{\beta}_{\rm b}$, by averaging the velocities of individual particles in that patch. We than compute the local Doppler factor $\delta=[\Gamma_{\rm b}(1-\hat{n}\cdot \boldsymbol{{\beta}}_{\rm b})]^{-1}$, which, for patch $(i,j)$, we call $\delta_{ij}$. Here, $\Gamma_{\rm b}=1/(1-\beta_{\rm b}^2)^{1/2}$ and $\hat{n}$ identifies the line of sight. We then weigh the local synchrotron flux $F_{ij}(\chi=1)$ by different powers $s$ of the  Doppler factor $\delta_{ij}$, as indicated by the legend, and finally sum over the patches, $\Sigma_{i,j}F_{ij}(\chi=1)\delta_{ij}^s$. Positive powers ($s>0$) artificially enhance the role of Doppler boosting, while negative powers tend to remove Doppler effects (i.e., they approach the case of emission computed in the fluid comoving frame). The lines for $s=0$ are the same as in \figg{syncLC}, i.e., they show the lightcurves extracted from our simulations, that self-consistently include Doppler effects. 

For $\alpha=0$, we find that the double-peaked nature of the lightcurve is emphasized even more with $s\geq1$, suggesting that Doppler effects have a key role in shaping the lightcurve. In fact, the same  double-peaked structure appears also for $\alpha=0.1$ if $s\geq1$. In contrast, $\alpha=0.3$ lightcurves peak at superior conjunction regardless of $s$. We then regard $\alpha=0.1$ as a borderline case (with a lightcurve ``almost'' double-peaked), and conclude that $\alpha\lesssim 0.1$ orientations are generally conducive to double-peaked lightcurves.


\section{Summary and Discussion}
We have performed global fully-kinetic particle-in-cell simulations of the intrabinary shock in spider pulsars. We assumed that the pulsar wind consists of plane-parallel stripes of alternating polarity and that the shock wraps around the companion star. We find that pulsar wind particles are efficiently accelerated at the shock via shock-driven reconnection. We extract first-principles synchrotron spectra and lightcurves and we find that: (1) the synchrotron spectrum
is nearly flat ($F_\nu\propto {\rm const}$); (2) when the pulsar spin axis is nearly aligned with the orbital angular momentum, the light curve displays two peaks, just before and after the pulsar eclipse (superior conjunction of the pulsar), separated in phase by  $\sim 0.8\, {\rm rad}$ (equivalently, $\sim 0.13$ in normalized phase); (3) the peak flux exceeds the one at inferior conjunction by a factor of ten. We also demonstrate that the double-peaked shape of the synchrotron lightcurve comes from Doppler boosting in the post-shock flow, as suggested by \cite{romani_sanchez_2016} and \cite{wadiasingh_2017}. All these properties are in good agreement with X-ray observations of spider pulsars. 

Our simulations reproduce, within one order of magnitude, the hierarchy of length scales expected in spider pulsars. However, we employ smaller values of $\gamma_0$ and $\sigma$ than expected in pulsar winds. While we have tested that a change in $\gamma_0$ only leads to a shift in the overall energy scale, upcoming simulations with larger $\sigma$ will be needed to clarify whether $\sigma=10$ is already in the asymptotic $\sigma\gg1$ limit, as local PIC simulations suggest \citep{sironi_spitkovsky_2011}. One may wonder, within our model, whether realistic values of $\gamma_0\sigma$ are such that the resulting emission at $\sim\omega_{\rm c}$ falls in the X-ray band.
Conservation of energy along the  pulsar wind streamlines implies that $\gamma_0 (1+\sigma) \kappa=\omega_{\rm LC}/2\,\Omega$, where $\omega_{\rm LC}=eB_{\rm LC}/mc$ is the 
cyclotron frequency at the light cylinder radius. We extrapolate the field from the pulsar surface to the light cylinder radius with a dipolar scaling, $B_{\rm LC} \sim B_{\rm P} (R_{\rm NS} / R_{\rm LC})^3$ (here, $R_{\rm NS}\sim 10\,{\rm km}$ is the neutron star radius). We assume that the post-shock field is of the same order as the pre-shock field $B_0 \sim B_{\rm LC} (R_{\rm LC}/R_{\rm TS})$. The characteristic synchrotron photon energy will be 
\begin{eqnarray}
\hbar\omega_{\rm c}&=&\hbar(\gamma_0\sigma)^2 \frac{eB_0}{mc}\\&\simeq& 0.2\left( \frac{10^4}{\kappa} \right )^2 \left ( \frac{10^{11}\,\mathrm{cm}}{{R_{\rm TS}}} \right ) \left ( \frac{B_{\rm P}}{10^9\,\mathrm{G}} \right )^3 \left ( \frac{\Omega}{10^3\,\mathrm{s^{-1}}} \right )^6 {\rm keV}~. 
\label{eq:peak}
\end{eqnarray}
The peaks of our synchrotron spectra are located at $\sim 20\hbar\omega_{\rm c}$.
 Thus, for realistic parameters, our conclusions on lightcurve shape and spectral hardness below the peak can be promptly applied to X-ray observations. We also remark that the synchrotron peak frequency, via \eqq{peak}, may provide important constraints on the multiplicity $\kappa$ of pulsar winds, if both the pulsar period $P$ and its derivative $\dot{P}$ (and so, also $B_{\rm P}\propto\sqrt{P\dot{P}}$) are measured.

We conclude with a few caveats. First, we have employed 2D simulations, and we defer to future work an assessment of 3D effects (e.g., for cases with $\alpha\neq 0$, in 2D field lines artificially accumulate ahead of the companion). Second, we have neglected the orbital motion of the system, which has been invoked to explain asymmetries in the light curve, with the two peaks having different heights \citep{kandel_21}. 
Third, we have ignored radiative cooling losses in the particle equation of motion. Depending on parameters, the post-shock flow may be slow- or fast-cooling \citep{wadiasingh_2017}, so our results are  applicable only to the slow-cooling cases. Finally, it will be desirable to extrapolate our results to larger  values of $R_{\rm curv}/\lambda$, as we will do in an upcoming work.

\acknowledgements
J.C. thanks the LSSTC Data Science Fellowship Program, which is funded by LSSTC, NSF Cybertraining Grant 1829740, the Brinson Foundation, and the Moore Foundation; his participation in the program has benefited this work. 
L.S. acknowledges support from the Cottrell Scholars Award, NASA 80NSSC20K1556, NSF PHY-1903412, DoE DE-SC0021254 and NSF AST-2108201.
The authors thank S. Bogdanov, V. Bosch-Ramon, A.K. Harding, B.D. Metzger, R.W. Romani, and Z. Wadiasingh for helpful and insightful comments.  


\bibliographystyle{aasjournal}
\bibliography{spiders_master,spider2}

\begin{thebibliography}{}
\expandafter\ifx\csname natexlab\endcsname\relax\def\natexlab#1{#1}\fi
\providecommand{\url}[1]{\href{#1}{#1}}
\providecommand{\dodoi}[1]{doi:~\href{http://doi.org/#1}{\nolinkurl{#1}}}
\providecommand{\doeprint}[1]{\href{http://ascl.net/#1}{\nolinkurl{http://ascl.net/#1}}}
\providecommand{\doarXiv}[1]{\href{https://arxiv.org/abs/#1}{\nolinkurl{https://arxiv.org/abs/#1}}}

\bibitem[{{Alpar} {et~al.}(1982){Alpar}, {Cheng}, {Ruderman}, \&
  {Shaham}}]{alpar_1982}
{Alpar}, M.~A., {Cheng}, A.~F., {Ruderman}, M.~A., \& {Shaham}, J. 1982, \nat,
  300, 728, \dodoi{10.1038/300728a0}

\bibitem[{{Arons} \& {Tavani}(1993)}]{arons_tavani_1993}
{Arons}, J., \& {Tavani}, M. 1993, \apj, 403, 249, \dodoi{10.1086/172198}

\bibitem[{{Bogdanov} {et~al.}(2011){Bogdanov}, {Archibald}, {Hessels}, {Kaspi},
  {Lorimer}, {McLaughlin}, {Ransom}, \& {Stairs}}]{bogdanov_2011}
{Bogdanov}, S., {Archibald}, A.~M., {Hessels}, J. W.~T., {et~al.} 2011, \apj,
  742, 97, \dodoi{10.1088/0004-637X/742/2/97}

\bibitem[{{Bogdanov} {et~al.}(2021){Bogdanov}, {Bahramian}, {Heinke}, {Freire},
  {Hessels}, {Ransom}, \& {Stairs}}]{bogdanov_21}
{Bogdanov}, S., {Bahramian}, A., {Heinke}, C.~O., {et~al.} 2021, \apj, 912,
  124, \dodoi{10.3847/1538-4357/abee78}

\bibitem[{{Bogdanov} {et~al.}(2005){Bogdanov}, {Grindlay}, \& {van den
  Berg}}]{bogdanov_grindlay_vandenberg_2005}
{Bogdanov}, S., {Grindlay}, J.~E., \& {van den Berg}, M. 2005, \apj, 630, 1029,
  \dodoi{10.1086/432249}

\bibitem[{{Bogovalov}(1999)}]{bogo_99}
{Bogovalov}, S.~V. 1999, \aap, 349, 1017.
\newblock \doarXiv{astro-ph/9907051}

\bibitem[{{Bogovalov} {et~al.}(2019){Bogovalov}, {Khangulyan}, {Koldoba},
  {Ustyugova}, \& {Aharonian}}]{bogo_19}
{Bogovalov}, S.~V., {Khangulyan}, D., {Koldoba}, A., {Ustyugova}, G.~V., \&
  {Aharonian}, F. 2019, \mnras, 490, 3601, \dodoi{10.1093/mnras/stz2815}

\bibitem[{{Bogovalov} {et~al.}(2012){Bogovalov}, {Khangulyan}, {Koldoba},
  {Ustyugova}, \& {Aharonian}}]{bogo_12}
{Bogovalov}, S.~V., {Khangulyan}, D., {Koldoba}, A.~V., {Ustyugova}, G.~V., \&
  {Aharonian}, F.~A. 2012, \mnras, 419, 3426,
  \dodoi{10.1111/j.1365-2966.2011.19983.x}

\bibitem[{{Bogovalov} {et~al.}(2008){Bogovalov}, {Khangulyan}, {Koldoba},
  {Ustyugova}, \& {Aharonian}}]{bogo_08}
{Bogovalov}, S.~V., {Khangulyan}, D.~V., {Koldoba}, A.~V., {Ustyugova}, G.~V.,
  \& {Aharonian}, F.~A. 2008, \mnras, 387, 63,
  \dodoi{10.1111/j.1365-2966.2008.13226.x}

\bibitem[{{Bosch-Ramon} {et~al.}(2012){Bosch-Ramon}, {Barkov}, {Khangulyan}, \&
  {Perucho}}]{boschramon_2012}
{Bosch-Ramon}, V., {Barkov}, M.~V., {Khangulyan}, D., \& {Perucho}, M. 2012,
  \aap, 544, A59, \dodoi{10.1051/0004-6361/201219251}

\bibitem[{{Bosch-Ramon} {et~al.}(2015){Bosch-Ramon}, {Barkov}, \&
  {Perucho}}]{boschramon_barkov_perucho_2015}
{Bosch-Ramon}, V., {Barkov}, M.~V., \& {Perucho}, M. 2015, \aap, 577, A89,
  \dodoi{10.1051/0004-6361/201425228}

\bibitem[{{Buneman}(1993)}]{buneman_1993}
{Buneman}, O. 1993, {in ``Computer Space Plasma Physics,'' Terra Scientific,
  Tokyo, 67}

\bibitem[{{Burrows} {et~al.}(1993){Burrows}, {Hubbard}, {Saumon}, \&
  {Lunine}}]{burrows_1993}
{Burrows}, A., {Hubbard}, W.~B., {Saumon}, D., \& {Lunine}, J.~I. 1993, \apj,
  406, 158, \dodoi{10.1086/172427}

\bibitem[{{Cerutti} \& {Philippov}(2017)}]{cerutti_philippov_2017}
{Cerutti}, B., \& {Philippov}, A.~A. 2017, \aap, 607, A134,
  \dodoi{10.1051/0004-6361/201731680}

\bibitem[{{Cerutti} {et~al.}(2020){Cerutti}, {Philippov}, \&
  {Dubus}}]{cerutti_philippov_dubus_2020}
{Cerutti}, B., {Philippov}, A.~A., \& {Dubus}, G. 2020, \aap, 642, A204,
  \dodoi{10.1051/0004-6361/202038618}

\bibitem[{{Ginzburg} \& {Quataert}(2020)}]{ginzburg_quataert_2020}
{Ginzburg}, S., \& {Quataert}, E. 2020, \mnras, 495, 3656,
  \dodoi{10.1093/mnras/staa1304}

\bibitem[{{Goldreich} \& {Julian}(1969)}]{goldreich_julian_1969}
{Goldreich}, P., \& {Julian}, W.~H. 1969, \apj, 157, 869,
  \dodoi{10.1086/150119}

\bibitem[{{Harding} \& {Gaisser}(1990)}]{harding_gaisser_1990}
{Harding}, A.~K., \& {Gaisser}, T.~K. 1990, \apj, 358, 561,
  \dodoi{10.1086/169009}

\bibitem[{{Huang} {et~al.}(2012){Huang}, {Kong}, {Takata}, {Hui}, {Lin}, \&
  {Cheng}}]{huang_2012}
{Huang}, R.~H.~H., {Kong}, A.~K.~H., {Takata}, J., {et~al.} 2012, \apj, 760,
  92, \dodoi{10.1088/0004-637X/760/1/92}

\bibitem[{{Huber} {et~al.}(2021){Huber}, {Kissmann}, {Reimer}, \&
  {Reimer}}]{huber_2021}
{Huber}, D., {Kissmann}, R., {Reimer}, A., \& {Reimer}, O. 2021, \aap, 646,
  A91, \dodoi{10.1051/0004-6361/202039277}

\bibitem[{{Kandel} {et~al.}(2019){Kandel}, {Romani}, \&
  {An}}]{kandel_romani_an_2019}
{Kandel}, D., {Romani}, R.~W., \& {An}, H. 2019, \apj, 879, 73,
  \dodoi{10.3847/1538-4357/ab24d9}

\bibitem[{{Kandel} {et~al.}(2021){Kandel}, {Romani}, \& {An}}]{kandel_21}
---. 2021, \apjl, 917, L13, \dodoi{10.3847/2041-8213/ac15f7}

\bibitem[{{Lamberts} {et~al.}(2013){Lamberts}, {Fromang}, {Dubus}, \&
  {Teyssier}}]{lamberts_2013}
{Lamberts}, A., {Fromang}, S., {Dubus}, G., \& {Teyssier}, R. 2013, \aap, 560,
  A79, \dodoi{10.1051/0004-6361/201322266}

\bibitem[{{Lu} {et~al.}(2021){Lu}, {Guo}, {Kilian}, {Li}, {Huang}, \&
  {Liang}}]{lu_2021}
{Lu}, Y., {Guo}, F., {Kilian}, P., {et~al.} 2021, \apj, 908, 147,
  \dodoi{10.3847/1538-4357/abd406}

\bibitem[{{Manchester}(2017)}]{manchester_2017}
{Manchester}, R.~N. 2017, Journal of Astrophysics and Astronomy, 38, 42,
  \dodoi{10.1007/s12036-017-9469-2}

\bibitem[{{P{\'e}tri} \& {Lyubarsky}(2007)}]{petri_lyubarsky_2007}
{P{\'e}tri}, J., \& {Lyubarsky}, Y. 2007, \aap, 473, 683,
  \dodoi{10.1051/0004-6361:20066981}

\bibitem[{{Philippov} {et~al.}(2020){Philippov}, {Timokhin}, \&
  {Spitkovsky}}]{philippov_timokhin_spitkovsky_2020}
{Philippov}, A., {Timokhin}, A., \& {Spitkovsky}, A. 2020, \prl, 124, 245101,
  \dodoi{10.1103/PhysRevLett.124.245101}

\bibitem[{{Phinney} {et~al.}(1988){Phinney}, {Evans}, {Blandford}, \&
  {Kulkarni}}]{phinney_1988}
{Phinney}, E.~S., {Evans}, C.~R., {Blandford}, R.~D., \& {Kulkarni}, S.~R.
  1988, \nat, 333, 832, \dodoi{10.1038/333832a0}

\bibitem[{{Roberts}(2013)}]{roberts_2013}
{Roberts}, M. S.~E. 2013, in Neutron Stars and Pulsars: Challenges and
  Opportunities after 80 years, ed. J.~{van Leeuwen}, Vol. 291, 127--132,
  \dodoi{10.1017/S174392131202337X}

\bibitem[{{Roberts} {et~al.}(2014){Roberts}, {Mclaughlin}, {Gentile}, {Aliu},
  {Hessels}, {Ransom}, \& {Ray}}]{roberts_2014}
{Roberts}, M.~S.~E., {Mclaughlin}, M.~A., {Gentile}, P., {et~al.} 2014,
  Astronomische Nachrichten, 335, 313, \dodoi{10.1002/asna.201312038}

\bibitem[{{Roberts} {et~al.}(2015){Roberts}, {McLaughlin}, {Gentile}, {Ray},
  {Ransom}, \& {Hessels}}]{roberts_2015}
{Roberts}, M. S.~E., {McLaughlin}, M.~A., {Gentile}, P.~A., {et~al.} 2015,
  arXiv e-prints, arXiv:1502.07208.
\newblock \doarXiv{1502.07208}

\bibitem[{{Romani} \& {Sanchez}(2016)}]{romani_sanchez_2016}
{Romani}, R.~W., \& {Sanchez}, N. 2016, \apj, 828, 7,
  \dodoi{10.3847/0004-637X/828/1/7}

\bibitem[{{Sanchez} \& {Romani}(2017)}]{sanchez_romani_2017}
{Sanchez}, N., \& {Romani}, R.~W. 2017, \apj, 845, 42,
  \dodoi{10.3847/1538-4357/aa7a02}

\bibitem[{{Sironi} \& {Spitkovsky}(2011)}]{sironi_spitkovsky_2011}
{Sironi}, L., \& {Spitkovsky}, A. 2011, \apj, 741, 39,
  \dodoi{10.1088/0004-637X/741/1/39}

\bibitem[{{Sironi} {et~al.}(2013){Sironi}, {Spitkovsky}, \&
  {Arons}}]{sironi_13}
{Sironi}, L., {Spitkovsky}, A., \& {Arons}, J. 2013, \apj, 771, 54,
  \dodoi{10.1088/0004-637X/771/1/54}

\bibitem[{{Spitkovsky}(2005)}]{spitkovsky_2005}
{Spitkovsky}, A. 2005, in AIP Conf. Ser., Vol. 801, Astrophysical Sources of
  High Energy Particles and Radiation, ed. {T.~Bulik, B.~Rudak, \&
  G.~Madejski}, 345, \dodoi{10.1063/1.2141897}

\bibitem[{{Timokhin} \& {Arons}(2013)}]{timokin_12}
{Timokhin}, A.~N., \& {Arons}, J. 2013, \mnras, 429, 20,
  \dodoi{10.1093/mnras/sts298}

\bibitem[{{Timokhin} \& {Harding}(2015)}]{timokhin_harding_2015}
{Timokhin}, A.~N., \& {Harding}, A.~K. 2015, \apj, 810, 144,
  \dodoi{10.1088/0004-637X/810/2/144}

\bibitem[{{van der Merwe} {et~al.}(2020){van der Merwe}, {Wadiasingh},
  {Venter}, {Harding}, \& {Baring}}]{vandermerwe_2020}
{van der Merwe}, C.~J.~T., {Wadiasingh}, Z., {Venter}, C., {Harding}, A.~K., \&
  {Baring}, M.~G. 2020, \apj, 904, 91, \dodoi{10.3847/1538-4357/abbdfb}

\bibitem[{{Wadiasingh} {et~al.}(2017){Wadiasingh}, {Harding}, {Venter},
  {B{\"o}ttcher}, \& {Baring}}]{wadiasingh_2017}
{Wadiasingh}, Z., {Harding}, A.~K., {Venter}, C., {B{\"o}ttcher}, M., \&
  {Baring}, M.~G. 2017, \apj, 839, 80, \dodoi{10.3847/1538-4357/aa69bf}

\bibitem[{{Wadiasingh} {et~al.}(2018){Wadiasingh}, {Venter}, {Harding},
  {B{\"o}ttcher}, \& {Kilian}}]{wadiasingh_2018}
{Wadiasingh}, Z., {Venter}, C., {Harding}, A.~K., {B{\"o}ttcher}, M., \&
  {Kilian}, P. 2018, \apj, 869, 120, \dodoi{10.3847/1538-4357/aaed43}

\end{thebibliography}

\end{document}